\documentclass[aps,pre,twocolumn,groupedaddress]{revtex4}

\usepackage{graphicx}

\begin{document}

\renewcommand{\vec}[1]{{\mathbf #1}}

\title{Information transfer through disordered media by diffuse waves}
\author{S.E. Skipetrov}
\email[]{Sergey.Skipetrov@grenoble.cnrs.fr}
\affiliation{Laboratoire de Physique et Mod\'elisation des Milieux Condens\'es,\\
CNRS, 38042 Grenoble, France}

\date{\today}

\begin{abstract}
We consider the information content $h$ of a scalar multiple-scattered, diffuse
wave field $\psi(\vec{r})$
and the information capacity $C$ of a communication channel that employs
diffuse waves
to transfer the information through a disordered medium.
Both $h$ and $C$ are shown to be directly related to
the mesoscopic correlations between the values of $\psi(\vec{r})$ at different
positions $\vec{r}$ in space, arising due to the coherent nature of the wave.
For the particular case of a communication channel between two identical linear
arrays of $n \gg 1$ equally-spaced transmitters/receivers (receiver spacing $a$),
we show that the average capacity
$\left< C \right> \propto n$ and obtain explicit analytic expressions
for $\left< C \right>/n$ in the limit
of $n \rightarrow \infty$ and $k \ell \rightarrow \infty$,
where $k= 2\pi/ \lambda$, $\lambda$ is the wavelength,
and $\ell$ is the mean free path.
Modification of the above results in the case of finite but large $n$ and
$k \ell$ is discussed as well.
If the signal to noise ratio $S/N$
exceeds some critical value $(S/N)_{\mathrm{c}}$,
$\left< C \right>/n$ is a non-monotonic function of $a$, exhibiting maxima
at $ka = m \pi$ ($m = 1, 2, \ldots$). For smaller
$S/N$, $ka = m \pi$ correspond to local minima, while the absolute maximum
of $\left< C \right>/n$ is reached at some $ka \sim (S/N)^{1/2} < \pi$.
We define the maximum average information capacity
$\left< C \right>_{\mathrm{max}}$
as $\left< C \right>$ maximized
over the receiver spacing $a$ and the optimal normalized receiver spacing
$(ka)_{\mathrm{opt}}$ as the spacing
maximizing $\left< C \right>$.
Both $\left< C \right>_{\mathrm{max}}/n$ and $(ka)_{\mathrm{opt}}$ scale
as $(S/N)^{1/2}$ for $S/N < (S/N)_{\mathrm{c}}$, while
$(ka)_{\mathrm{opt}} = m \pi$ and
$\left< C \right>_{\mathrm{max}}/n \sim \log(S/N)$ for
$S/N > (S/N)_{\mathrm{c}}$.
\end{abstract}

\pacs{42.25.Dd, 84.40.Ua, 89.70.+c}

\maketitle


\section{Introduction}
\label{intro}

Transport of coherent waves in
disordered media have been extensively studied during the
last decades \cite{alt91,berk94,datta95,lagendijk96,rossum99,sebbah01}.
Remarkably similar, diffusion behavior of multiple-scattered
electronic wave functions at low temperatures \cite{alt91,berk94,datta95},
coherent electromagnetic \cite{rossum99}
(optical \cite{lagendijk96} and microwave \cite{sebbah00}),
acoustic \cite{derode01}, and elastic \cite{treg02a}
waves has permitted impressive advances of the field \cite{sebbah01}.
Recently, multiple-scattered seismic waves in the
Earth crust have been demonstrated to behave in a similar way \cite{henino01,treg02b}.
In the context of these studies, the main quantities of interest
are the transport coefficients of disordered samples, such as, e.g., the transmission
coefficient $T$ or the conductance $g$. The average values,
fluctuations, full probability distributions, angular, spatial, and
temporal correlation functions of $T$ and $g$ have been studied both theoretically and
experimentally (see, e.g., Ref.\ \onlinecite{rossum99} for a review). These
quantities are, without any doubt, very important, since they describe the
transport of wave \textit{energy} through a disordered sample and hence can be measured
experimentally. On the other hand, in practical applications one rarely uses
multiple-scattered waves with a primary purpose of energy transmission. Much more
often, waves are used to transfer the \textit{information}.
Readily available examples are microwave communications (portable telephony)
in cities, indoor wireless local-area networks in building with complex structure,
and underwater acoustic communication systems.
In fact, one of the main motivations to study the multiple scattering of waves in
disordered media is their possible use for transmission or processing of
information in electronic devices or wireless communications.

In the same way as the properties of a disordered sample with respect to
transmission of energy are described by the transmission coefficient
$T$ or conductance $g$ (depending on the details of the specific experiment),
its properties with respect to transmission of information are
characterized by the \textit{information capacity} $C$.
The latter gives the maximum rate of error-free information transfer
through a disordered sample using a given type of waves (acoustic, electromagnetic,
etc.) and a given transmitter/receiver configuration (see
Refs.\ \onlinecite{shannon48,cover91,gray90} for a more rigorous definition of $C$).
Information capacity of communication channels in disordered media has recently
received considerable attention
\cite{fosc98,mous00,andrews01,simon01,sengupta01,skip02,mous02}.
The most interesting and important
result concerns a communication system consisting of multiple
transmitters and receivers \cite{fosc98,mous00,simon01,sengupta01,skip02,mous02}:
it has been found that the capacity
of such a communication system scales linearly with the number of receivers $n$
(as long as the number of transmitters $m$ is of the same order). The authors of
Refs.\ \onlinecite{mous00,sengupta01,skip02,mous02} have also studied the effect of
correlations between the signals received by different receivers on $C$.
More specifically, the communication channel between $m$ transmitters and
$n$ receivers is described by a $n \times m$ Green matrix $G$
($G_{\alpha i}$ is the signal at the receiver $\alpha$ due to a unit
signal emitted by the transmitter $i$). Due to the multiple scattering of waves, in
a disordered medium the entries of $G$ are random variables
with certain correlations
between them. These correlations are often termed ``mesoscopic'' \cite{rossum99}
because
they originate from the fact that the phase coherence length of the considered wave
exceeds the length of the path that the wave travels inside the disordered medium.
Mesoscopic correlations complicate significantly the
theoretical calculation of capacity which otherwise (i.e., for uncorrelated
$G_{\alpha i}$) is relatively straightforward \cite{fosc98,sengupta01,skip02}.
Although it follows from Refs.\ \onlinecite{mous00} and \onlinecite{skip02} that
nonzero correlations between $G_{\alpha i}$ reduce the information capacity
of the communication channel under particular conditions considered in that papers,
no systematic study of the role of mesoscopic correlations in the context of
information transfer is available at the time of writing.
Mesoscopic
correlations have been extensively studied during the last decade
(see, e.g., Refs.\ \onlinecite{berk94,rossum99,sebbah00} and references therein),
but their role in the context of the
information transfer has been completely ignored until
very recently \cite{mous00,sengupta01,skip02}.
Meanwhile, understanding the relation between mesoscopic
correlations and information capacity has not only the practical importance
but also a fundamental significance, since the
correlations are usually affected by the symmetries of the problem
(translational symmetry, time-reversal symmetry, etc.) and it might be interesting
to study the influence of these symmetries on the information-theoretic quantities,
such as the information capacity.

\begin{figure}
\includegraphics[width=8cm]{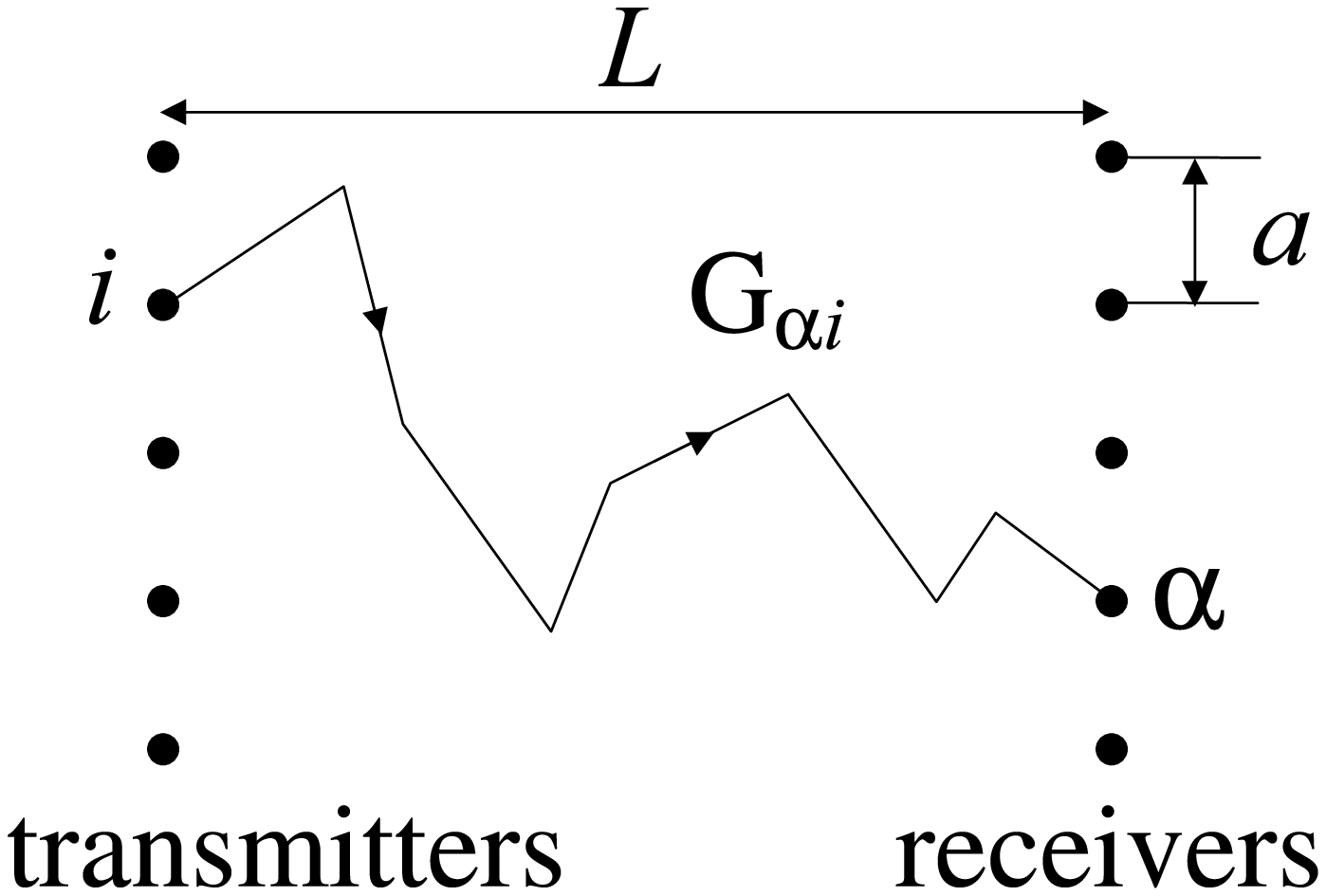}
\caption{\label{fig1} We consider the transfer of information between identical
arrays of $n$ transmitters/receivers placed in a disordered medium.
We denote the transmitter/receiver spacing by $a$ and the distance between the
arrays by $L$. A wave emitted by some transmitter $i$ experiences multiple scattering
in the disordered medium before reaching the receiver $\alpha$.
The signal measured by the receiver $\alpha$ due
to the transmitter $i$ is given by the Green function $G_{\alpha i}$.
The Green functions $G_{\alpha i}$ ($\alpha, i = 1, \ldots, n$) form a
Green matrix $G$.}
\end{figure}

The purpose of the present paper is to consider the multiple-scattered wave
field from the point of view of the information theory, to quantify its information
content, and, finally, to
provide a comprehensive study of the
information capacity of a communication channel in a disordered medium with a proper
account for mesoscopic correlations arising from the coherent nature of
multiple-scattered waves that carry the information. To be specific, we limit our
consideration to identical linear arrays of $n$ equally-spaced
transmitters/receivers (see Fig. \ref{fig1}).
Such a geometry is common for microwave \cite{sebbah00} and acoustic \cite{derode01}
experiments with multiple-scattered waves in disordered media
\footnote{To reduce the price of the experimental setup, one often uses
a single mobile receiver to scan the multiple scattered speckle pattern.
As long as the speckle pattern is stationary, this is equivalent to performing
a simultaneous measurement with multiple receivers at different positions.}.
Besides, it is also widely considered in connection with the
time reversed acoustics \cite{fink97} and
the BLAST architecture for efficient communication over fading wireless
channels \cite{blast}. We note that the time-reversal techniques
seems to be very promising for wireless communications in disordered and/or
chaotic environments \cite{hein02}.

The paper is organized as follows. We start by a study of the
information content of the multiple-scattered wave field in Sec.\ \ref{entropysec}.
The information content of a random field is quantified by its differential entropy.
We show that mesoscopic correlations reduce the differential entropy, reducing the
information content of the multiple-scattered wave field and leading, in
practice, to a
smaller volume of computer memory required for its storage.
After this somewhat introductory part of the paper, in Sec.\ \ref{capacitysec}
we study the information
capacity $C$ of a communication channel between two identical linear arrays of
$n$ transmitters/receivers in a disordered medium (see Fig.\ \ref{fig1}).
Under certain approximations, we obtain analytic expressions for the average
information capacity $\left< C \right>$. The latter expressions provide a reasonable
estimate of $\left< C \right>$
in real experimental situations, although one can do better
using numerical methods. Next, we show that mesoscopic correlations reduce the
information capacity (as found in Ref.\ \onlinecite{mous00}) only if the signal
power is strong enough. If the signal is weak, correlations play a positive role
and allow a higher capacity as compared to that in the absence of correlations.
Finally, we define the maximum capacity
as a capacity maximized over the transmitter/receiver spacing and
the optimal transmitter/receiver spacing as the spacing maximizing the capacity.
Simple analytic expressions are found for these two quantities.
We summarize the main results of the paper in Sec.\ \ref{concl}.
Derivations of some important equations used in the main text but not essential
for its understanding are provided in the Appendices \ref{appA} and \ref{appB}.

\section{Information content of the multiple-scattered wave field}
\label{entropysec}

\begin{figure}
\includegraphics[width=8cm]{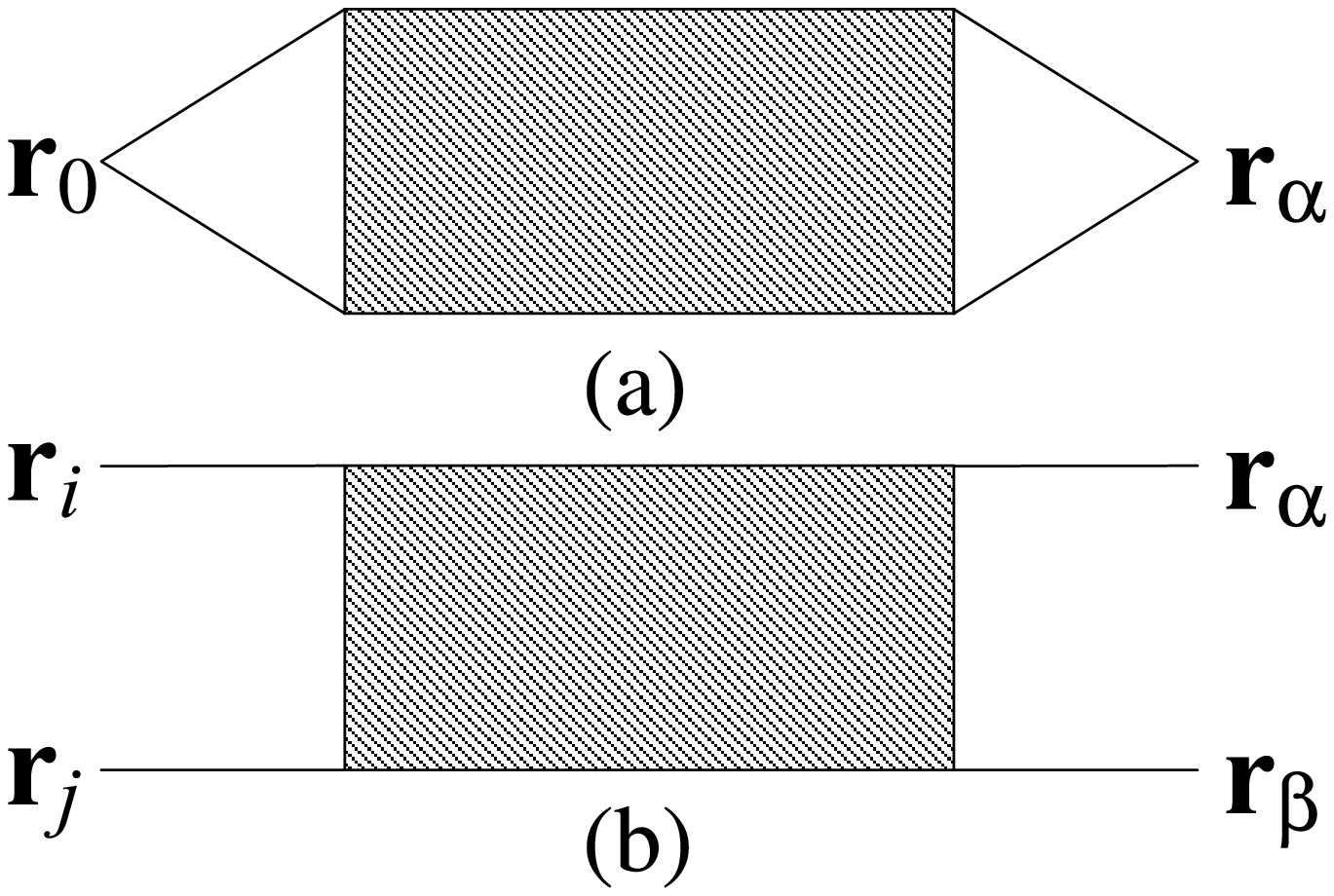}
\caption{\label{fig2} (a) The diagram contributing to the average intensity
$\left< I(\vec{r}_{\alpha}) \right>$ for a point source of radiation at
$\vec{r}_0$. The shaded rectangle is the ladder propagator,
the solid lines denote the retarded and advanced average Green functions.
(b) The diagram for the correlation function
of the Green functions $\left< G_{\alpha i} G_{\beta j}^* \right>$.}
\end{figure}

We start our analysis by considering a problem which is somewhat simpler
than the one depicted in Fig.\ \ref{fig1}. Namely, we assume that there is
only one transmitter (located, say, at $\vec{r}_0$) that emits a monochromatic
scalar spherical wave, and analyze the information content of the multiple-scattered
wave field $\psi(\vec{r})$. As a consequence of multiple scattering, $\psi(\vec{r})$
is a random function of position $\vec{r}$ (``speckle pattern'').
Since $\psi(\vec{r})$ is a complex continuous random field, it contains
an infinite amount of information or, in other words, an infinite volume of
computer memory would be required to store its values at all $\vec{r}$
with absolute accuracy. In reality, however, one usually measures $\psi(\vec{r})$
at some finite number $n$ of positions $\vec{r}_{\alpha}$
($\alpha = 1, \ldots, n$). In the following, we identify the information content
of such a measurement with the information content of $\psi(\vec{r})$, keeping
in mind that this is strictly true only for $n \rightarrow \infty$.
A measurement of $\psi$ in $n$ points $\vec{r}_{\alpha}$
results in a random complex vector $\vec{y} = \left\{ y_{\alpha} \right\}$, where
$y_{\alpha} = \psi(\vec{r}_{\alpha})/\left< I \right>^{1/2}$ and
we normalize the field $\psi(\vec{r})$ by the square root of the average
intenisty $\left< I \right> = \left< \left| \psi(\vec{r}_{\alpha}) \right|^2 \right>$
given by the diagram (a) of Fig.\ \ref{fig2} and
assumed to be the same for all $n$ receivers. From here on the angular
brackets denote averaging over disorder.
If $p(\vec{y})$ is a probability density function of
the random vector $\vec{y}$, the \textit{differential entropy} of $\vec{y}$ is
defined as \cite{shannon48,cover91}
\begin{eqnarray}
h(\vec{y}) = -\int p(\vec{y}) \log p(\vec{y}) \mathrm{d}^n \vec{y}.
\label{entropy}
\end{eqnarray}
The logarithm is to base $2$ and $h$ is hence
measured in \textit{bits}. The differential entropy $h$ is a measure of uncertainty of
the random vector $\vec{y}$ and reflects the average amount of information that one
receives when some value of $\vec{y}$ is observed.
The larger the differential entropy, the larger the
information content of the vector $\vec{y}$, and we will therefore use
$h$ to quantify the information content of $\vec{y}$ and, consequently,
of the multiple-scattered wave field $\psi(\vec{r})$.

Under conditions of strong multiple scattering and provided that
$\left| \vec{r}_{\alpha} - \vec{r}_0 \right| \gg \ell$ for all
$\alpha = 1, \ldots, n$ and that  $k \ell \gg 1$
(where $k = 2 \pi / \lambda$, $\lambda$ is the wavelength, and $\ell$ is the
mean free path), the wave field $\psi(\vec{r})$ can be considered
Gaussian to a good accuracy, and hence $\vec{y}$ is a circularly symmetric
complex Gaussian random vector described by a Gaussian probability density function
\begin{eqnarray}
p(\vec{y}) = \det(\pi K)^{-1} \exp\left(-\vec{y}^+ K^{-1} \vec{y} \right),
\label{gauss}
\end{eqnarray}
where $K$ is a covariance matrix:
$K_{\alpha \beta} = \left< y_{\alpha} y_{\beta}^* \right>$.
The integration in Eq.\ (\ref{entropy}) can be then
carried out analytically, yielding
\begin{eqnarray}
h(\vec{y}) = \log \det(\pi e K).
\label{entropy2}
\end{eqnarray}
It also can be shown \cite{cover91} that the Gaussian density function
(\ref{gauss}) maximizes the differential entropy for a given covariance matrix $K$.
Recalling that $K_{\alpha \beta}$ is the correlation function of the multiple-scattered
wave field, we can
readily compute $K$ in the ladder approximation [see the diagram (b) of
Fig.\ \ref{fig2} with $\vec{r}_i = \vec{r}_j = \vec{r}_0$
or Ref.\ \onlinecite{shapiro86}]:
\begin{eqnarray}
K_{\alpha \beta} =
\frac{\sin(k \Delta r_{\alpha \beta})}{k \Delta r_{\alpha \beta}}
\exp \left( -\frac{\Delta r_{\alpha \beta}}{2 \ell} \right),
\label{kij}
\end{eqnarray}
where $\Delta r_{\alpha \beta} = \left| \vec{r}_{\alpha} - \vec{r}_{\beta}
\right|$.
We now restrict ourselves to the case of measurement points
$\vec{r}_{\alpha}$ arranged in a line with a constant distance $a$ between
$\vec{r}_{\alpha}$ and $\vec{r}_{\alpha+1}$.
As we already mentioned in Sec.\ \ref{intro},
such a measurement geometry is common for both microwave \cite{sebbah00}
and acoustic \cite{derode01,fink97} experiments.
The covariance matrix $K$ is then Toeplitz:
$K_{\alpha \beta} = K_{\alpha-\beta}$, where
$K_{\gamma} = \sin[\gamma ka]/[\gamma ka]
\exp[-\left| \gamma \right| ka/(2 \ell)]$,
and hence in the limit $n \rightarrow \infty$ the density of its eigenvalues
tends to a limit, which is the spectrum of $\vec{y}$.
The differential entropy rate can be then expressed through the power spectral density
of $\vec{y}$ \cite{bot99}
\begin{eqnarray}
f(\mu) = \sum\limits_{\gamma=-\infty}^{\infty} K_{\gamma} \exp(i \gamma \mu).
\label{f}
\end{eqnarray}
The resulting $h(\vec{y})$
appears to scale linearly with $n$ and hence
it is convenient to consider the \textit{differential entropy rate} or
the differential entropy per receiver
\begin{eqnarray}
{\cal H}(\vec{y}) = \lim_{n \rightarrow \infty}
\frac{h(\vec{y})}{n} = \log(\pi e) + \frac{1}{2 \pi}
\int_0^{2 \pi} \log f(\mu) \mathrm{d} \mu.
\label{entropy3}
\end{eqnarray}

\begin{figure}
\includegraphics[width=8cm]{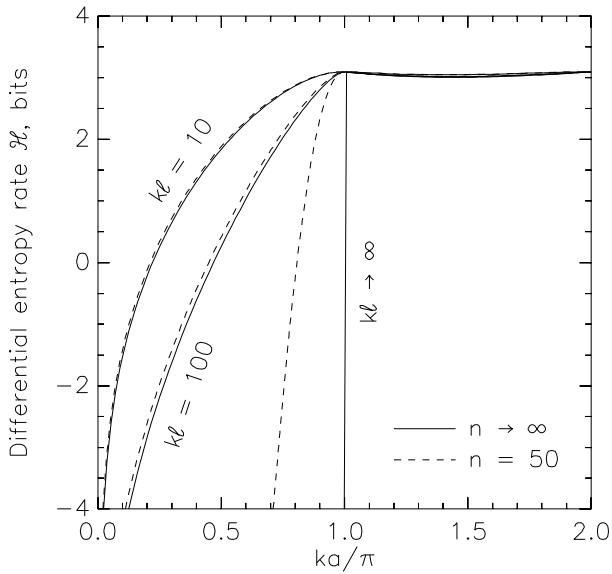}
\caption{\label{fig3} Differential entropy rate ${\cal H}$ as a function of
normalized receiver spacing for the
multiple-scattered wave field measured by a linear array of
$n \rightarrow \infty$ (solid lines) and $n = 50$ (dashed lines) receivers at
$k \ell = 10$, $100$, and $k \ell \rightarrow \infty$.}
\end{figure}

Equation (\ref{kij}) allows us to compute $f(\mu)$ explicitly:
\begin{eqnarray}
f(\mu) &=& 1 + \frac{1}{ka} \left\{ \arctan
\frac{\sin(ka + \mu)}{\exp \left[ ka/(2 k \ell) \right] - \cos(ka + \mu)}
\right.
\nonumber \\
&+& \left. \arctan
\frac{\sin(ka - \mu)}{\exp \left[ ka/(2 k \ell) \right] - \cos(ka - \mu)} \right\}.
\label{f2}
\end{eqnarray}
This expression simplifies greatly in the limit $k \ell \rightarrow \infty$:
\begin{eqnarray}
f(\mu) = \frac{\pi}{ka} \left( m_{+} + m_{-} + 1 \right),
\label{f3}
\end{eqnarray}
where $m_{\pm}$ denotes the greatest integer not larger than $(ka \pm \mu)/(2 \pi)$.
We show the differential entropy rate following from Eqs.\
(\ref{entropy3}--\ref{f3}) in Fig.\ \ref{fig3} by solid lines for
$k \ell = 10$, $100$, and $k \ell \rightarrow \infty$.
In order to demonstrate the convergence of the exact formula
(\ref{entropy2}) to the asymptotic Eq.\ (\ref{entropy3}), we also show the curves
following from Eq.\ (\ref{entropy2}) for $n = 50$
(dashed lines) and the same values of $k \ell$.
It is worthwhile to note that the entropy rate
${\cal H}$ reaches its maximum value $\log(\pi e)$
at $ka = m \pi$ (where $m$ is a positive
integer) and that for $ka > \pi$ the variations of ${\cal H}$ with $ka$ are extremely
weak. Exact analytic result for ${\cal H}$ can be obtained in the limit
$k \ell \rightarrow \infty$ by substituting $f(\mu)$ given by Eq.\ (\ref{f3}) into
Eq.\ (\ref{entropy3}) and performing the integration.
We find ${\cal H} \rightarrow -\infty$ for $ka < \pi$ and
${\cal H} = ka/\pi + \log[(\pi e/2)(\pi/ka)]$ for $\pi < ka < 2 \pi$.
At $ka > 2 \pi$, ${\cal H}$ shows only weak deviations from its maximum value
$\log(\pi e)$.

Let us now discuss the implication of Fig.\ \ref{fig3} for experimental measurements.
In an experiment,
the values of $y_{\alpha}$ cannot be measured with an absolute accuracy.
A $m$-bit quantization of $\mathrm{Re} y_{\alpha}$ and  $\mathrm{Im} y_{\alpha}$
with a quantization step $\Delta \sim 2^{-m}$ is a common procedure \footnote{Only real
quantities are, of course, measured in experiments. The signal has, however,
two degrees of freedom: the amplitude $A = [(\mathrm{Re} y_{\alpha})^2 +
(\mathrm{Im} y_{\alpha})^2]^{1/2}$ and the phase $\phi
= \arctan (\mathrm{Im} y_{\alpha} / \mathrm{Re} y_{\alpha})$.
Measuring $A$ and $\phi$ is equivalent to measuring the complex variable
$y_{\alpha}$.} and the measured quantized
$y_{\alpha}^{\,\prime}$ can take $\sim \Delta^{-2}$ discrete values.
The vector $\vec{y}^{\,\prime}=\left\{ y_{\alpha}^{\,\prime} \right\}$ can therefore
take $\sim \Delta^{-2n}$ different values.
One can show that
the entropy $H(\vec{y}^{\,\prime})$ of $\vec{y}^{\,\prime}$ is approximately
$h(\vec{y}) - 2 n \log \Delta$ for $\Delta \rightarrow 0$.
The number of bits required on average to describe a given component
$y_{\alpha}^{\,\prime}$ of the
random vector $\vec{y}^{\,\prime}$ is then $H(\vec{y}^{\,\prime})/n
\sim {\cal H}(\vec{y}) - 2 \log \Delta$.
It can be now easily seen that the smaller
differential entropy rate ${\cal H}(\vec{y})$ means that less bits will be required
in an experiment to record all the relevant information about the multiple-scattered
wave field. As follows from Fig.\ \ref{fig3}, at large receiver spacing $a$
($ka > \pi$), ${\cal H}$ is very close to its maximum value. At such a large
receiver spacing, the signals measured by different receivers are only weakly
correlated and hence each signal contains a relatively large amount of information.
One therefore needs a relatively large number of bits per receiver
$H(\vec{y}^{\,\prime})/n$ to record the speckle pattern $\psi(\vec{r})$
in this case.
In contrast, at small $a$ ($ka < \pi$), ${\cal H}$ decreases, since the
signals measured by different receivers become significantly correlated and hence
the information contained in each of the signals on average,
$H(\vec{y}^{\,\prime})/n$, is smaller than at $ka > \pi$.
If $k \ell$ or $n$ is finite, the value of ${\cal H}$ at $ka < \pi$ can be
small but remains finite.
If, in contrast, we take the limit of $k \ell \rightarrow \infty$ and
$n \rightarrow \infty$, ${\cal H} \rightarrow -\infty$ at $ka < \pi$.
This means that an infinitely small amount of information
is contained in the signal measured by a given receiver and is a direct consequence
of the fact that in the absence of exponential damping of correlation in Eq.\ (\ref{kij}),
the correlation range is infinite and the decrease of
$h(\vec{y})$ with $n$ is faster than linear.

\section{Information capacity of a communication channel in a disordered
medium}
\label{capacitysec}

\subsection{General definitions}

After having considered the information content of the multiple-scattered wave
field, we are now in a position to analyze the central
problem of the present paper. We consider a communication channel between
two identical linear arrays of $n$ equally-spaced
transmitters/receivers shown in
Fig.\ \ref{fig1}. The vector of emitted signals $\vec{x}$ and the vector of
received
signals $\vec{y}$ are related by $\vec{y} = G \vec{x} + \vec{z}$, where $G$ is
a $n \times n$ complex Green matrix ($G_{\alpha i}$ gives the signal measured
at the receiver $\alpha$ due to a unit signal emitted by the transmitter $i$),
and $\vec{z}$ is a noise vector. We consider scalar waves and
assume that the noises $z_{\alpha}$
at different receivers are statistically independent, normally distributed
random variables with power $N$: $\left< z_{\alpha} z_{\beta}^* \right> =
N \delta_{\alpha \beta}$.
Before defining the information capacity of such a communication channel,
we first remind the definitions of the \textit{conditional differential entropy}
\cite{cover91}
\begin{eqnarray}
h(\vec{y} | \vec{x}) = -\int\int p(\vec{x}, \vec{y}) \log p(\vec{y} | \vec{x})
\mathrm{d}^n \vec{x} \, \mathrm{d}^n \vec{y},
\label{cond}
\end{eqnarray}
and \textit{mutual information} between two random vectors $\vec{x}$ and $\vec{y}$
\cite{cover91}:
\begin{eqnarray}
{\cal I}(\vec{x}, \vec{y}) = h(\vec{y}) - h(\vec{y} | \vec{x}),
\label{mut}
\end{eqnarray}
where $p(\vec{x}, \vec{y})$ and $p(\vec{y} | \vec{x})$ are the joint and
conditional probability density functions of $\vec{x}$ and $\vec{y}$,
respectively. In our case, $\vec{y} = G \vec{x} + \vec{z}$, and
one finds $h(\vec{y} | \vec{x}) = h(\vec{z})$ and
\begin{eqnarray}
{\cal I}(\vec{x}, \vec{y}) = \log \det \left[ I_n + \frac{1}{N} G^+ Q G \right].
\label{mut2}
\end{eqnarray}
Here $I_n$ is the $n \times n$ unit matrix and
we assume that $\vec{x}$ is a circularly symmetric complex
Gaussian random vector with a covariance matrix $Q$:
$Q_{ij} = E \left[  x_i x_j^* \right]$,
where $E[\ldots]$ denotes the averaging over all possible emitted signals
$\vec{x}$ and should be contrasted from the disorder averaging that we denote
by the angular brackets.

If we assume that the Green matrix $G$ is known at both transmitters and receivers,
the Shannon \textit{information capacity} $C$ (or simply capacity for brevity)
of the communication channel
is found by maximizing the mutual information (\ref{mut}) over all possible
distributions $p(\vec{x})$ of emitted signals $\vec{x}$.
The fundamental importance of $C$ is that it gives the largest information transfer
rate $R$ that can be realized for a given information channel \textit{in principle}
with infinitely small
probability of error \cite{shannon48,cover91}. Although no general procedure exists
to realize $R = C$ in practice, in many real situations one can achieve
information transfer rates that are quite close to
$C$ \footnote{For example, the transmission rate used in the
telephony is about 90\% of the information capacity of a typical
telephone line \cite{cover91}.}.
Since the Gaussian distribution maximizes the differential
entropy $h$
(and hence the mutual information ${\cal I}$) at a given $Q$ (see Sec.\ \ref{entropysec}
and Ref.\ \onlinecite{cover91}), maximizing Eq.\ (\ref{mut}) over $p(\vec{x})$
amounts to maximize Eq.\ (\ref{mut2}) over $Q$.
In practice, however, $G$ is often known only at the receivers, but not at the
transmitters, where only statistical information about $G$ is available. In this case
the optimal matrix $Q$ is that maximizing the \textit{average} mutual
information $\left< {\cal I}(\vec{x}, \vec{y}) \right>$.
To accomplish the averaging, we need to specify the statistical
properties of the Green matrix $G$. In the considered case of strong multiple
scattering, provided that the distance $L$ between the arrays of transmitters
and receivers is much larger than the mean free path $\ell$ and that
$k \ell \gg 1$, $G_{\alpha i}$ is a circularly symmetric complex Gaussian random
variable with zero mean
and covariance given by the diagram (b) of Fig.\ \ref{fig2}:
$\left< G_{\alpha i} G_{\beta j}^* \right> =
\left< I \right> K_{\alpha \beta} K_{ij} $, where
$K_{\alpha \beta}$ is defined in Eq.\ (\ref{kij}) and
$\left< I \right> = \left< \left| G_{\alpha i} \right|^2 \right>$ is assumed
to be independent of $\alpha$ and $i$.
In the following we adopt the total emitted power constraint
$\mathrm{Tr} Q \leq n$ and introduce a normalized Green matrix
${\cal G} = G/(\left< I \right> n)^{1/2}$.
In the present paper we will only be interested in the average capacity
$\left< C \right>$, although it should be kept in mind that $C$ exhibits
random fluctuations as disorder is varied.
As follows from the above reasoning, the average capacity is
\begin{eqnarray}
\left< C \right> = \max_{Q}
\left< \log \det \left[ I_n + \frac{S}{N} {\cal G}^+ Q {\cal G} \right] \right>,
\label{cap0}
\end{eqnarray}
where $S/N$ plays the role of the signal to noise ratio and
$S = n \left< I \right>$ is the average power received by each receiver assuming
independent signals from transmitters.
When Eq.\ (\ref{cap0}) is applied to a real situation, it gives the maximum
amount of information that can be transferred through the considered communication
channel per second using a unit frequency bandwidth \cite{shannon48,cover91}.
The units of
$\left< C \right>$ are therefore bits per second per Hertz (or bps/Hz for
brevity).

\subsection{Average capacity at $n \gg 1$}

For a small number of transmitters/receivers $n \sim 1$, the averaging and maximization
over $Q$ in Eq.\ (\ref{cap0}) can be carried out by a numerical simulation
(see, e.g., Ref.\ \onlinecite{skip02} for $n=2$ and Ref.\ \onlinecite{mous02}
for $n = 2$ and $3$).
At large $n$ such an approach becomes inadequate. It appears, however, that
analytic methods can be applied to estimate the asymptotic behavior of
capacity for $n \gg 1$ (see Appendix \ref{appA} and
Refs.\ \onlinecite{mous00,sengupta99,sengupta01}).
In the limit of large $n$ the capacity per receiver is given by
\begin{eqnarray}
\frac{\left< C \right>}{n} &=&  \frac{1}{n} \sum\limits_{i=1}^{n}
\log \left[(S/N)^{-1/2} + \kappa_i q_i u \right]
\nonumber \\
&+& \frac{1}{n} \sum\limits_{i=1}^{n} \log \left[(S/N)^{-1/2} + \kappa_i v \right]
\nonumber \\
&-& u v/\ln 2 + \log(S/N),
\label{cap}
\end{eqnarray}
where $\kappa_i$ are the eigenvalues of the matrix $K$, while the
auxiliary variables $u$, $v$, and $p \leq n$ nonzero
eigenvalues $q_i$ of the matrix $Q$
are solutions of the following system of equations (see Appendix \ref{appA} for
derivations):
\begin{eqnarray}
u &=& \frac{1}{n} \sum\limits_{i=1}^n \frac{\kappa_i}{(S/N)^{-1/2} + \kappa_i v},
\label{u}
\\
v &=& \frac{1}{n} \sum\limits_{i=1}^n \frac{\kappa_i q_i}{(S/N)^{-1/2} +
\kappa_i q_i u},
\label{v}
\\
\phi &=& \frac{\kappa_i u}{(S/N)^{-1/2} + \kappa_i q_i u},
\;\;\; i = 1, \ldots, p,
\label{phi}
\\
n &=& \sum\limits_{i=1}^p q_i.
\label{power}
\end{eqnarray}
In principle, the above equations are sufficient for
the calculation of the average capacity at given $n \gg 1$, $S/N$, and $ka$.
It should be noted, however, that the total number of equations is
$p+3$ (with $p$ that can be as large as $n$),
and that the equations are nonlinear. Hence, the numerical solution requires
considerable computational resources at large $n$.
Besides, the interpretation of numerical results is known to be a rather
difficult task. Below we show that in the limit of $n \rightarrow \infty$,
Eqs.\ (\ref{cap})--(\ref{power}) can be significantly simplified and
even that simple \textit{analytic} expressions for $\left< C \right>$ can be obtained
in certain cases.

\subsection{Average capacity at $n \rightarrow \infty$}

As we already mentioned
in Sec.\ \ref{entropysec}, the covariance matrix $K$ is Toeplitz and hence
for $n \rightarrow \infty$ one can use the limit theorems known
for this class of matrices \cite{bot99}.
Of particular use for us is the fundamental eigenvalue distribution
theorem of Szeg\"{o} that states that if $K$ is a $n \times n$ Hermitian
Toeplitz matrix and $F(x)$ is some continuous function then
(under certain conditions fulfilled in our case)
\begin{eqnarray}
\lim_{n \rightarrow \infty} \frac{1}{n} \sum\limits_{i = 1}^{n}
F(\kappa_i) = \frac{1}{2 \pi} \int_0^{2 \pi}
F\left[ f(\mu) \right] \mathrm{d} \mu,
\label{szego}
\end{eqnarray}
where $f(\mu)$ is the power spectral density defined in Eq.\ (\ref{f}).
We now admit that the right-hand sides of
Eqs.\ (\ref{cap})--(\ref{v}) can be readily simplified using
Eq.\ (\ref{szego}). After some algebra this leads to the following set
of equations for the auxiliary variables $u$ and $v$:
\begin{eqnarray}
u &=& \frac{1}{2 \pi} \int_0^{2 \pi}
f(\mu) \left[ \left( \frac{S}{N} \right)^{-1/2} + v f(\mu) \right]^{-1}
\mathrm{d} \mu,
\label{u1}
\\
v &=& \frac{1}{u} \frac{1}{2 \pi} \int_{0\;\; f(\mu) > v/\sqrt{S/N}}^{2 \pi}
\nonumber \\
&&\left[ 1 - \left( \frac{S}{N} \right)^{-1/2} \frac{v}{f(\mu)} \right]
\mathrm{d} \mu,
\label{v1}
\end{eqnarray}
while the expression (\ref{cap}) for the average capacity per receiver becomes
\begin{eqnarray}
\frac{\left< C \right>}{n} &=&
\frac{1}{2 \pi} \int_{0\;\; f(\mu) > v/\sqrt{S/N}}^{2 \pi}
\log \left[ \frac{f(\mu)}{v} \right] \mathrm{d} \mu
\nonumber \\
&+&
\frac{1}{2 \pi} \int_0^{2 \pi}
\log \left[ \left( \frac{S}{N} \right)^{-1/2} + v f(\mu) \right] \mathrm{d} \mu
\nonumber \\
&+& \log \left( \frac{S}{N} \right) \left( 1 -
\frac{1}{4 \pi} \int_{0\;\; f(\mu) < v/\sqrt{S/N}}^{2 \pi}
\mathrm{d} \mu \right)
\nonumber \\
&-& \frac{u v}{\ln2},
\label{c1}
\end{eqnarray}
where the integral in Eq.\ (\ref{v1}) and the first integral in Eq.\ (\ref{c1})
are over the part of the interval $(0, 2 \pi)$ where
$f(\mu) > (S/N)^{-1/2} v$, while the last integral in Eq.\ (\ref{c1})
is over the part of the same interval where $f(\mu) < (S/N)^{-1/2} v$.
Once $f(\mu)$ is known, Eqs.\ (\ref{u1})--(\ref{c1}) allow one to calculate
the average capacity per receiver,
$\left< C \right>/n$, in the limit $n \rightarrow \infty$.

\subsubsection{Average capacity at $n \rightarrow \infty$ and
$k \ell \rightarrow \infty$}

\begin{figure}
\includegraphics[width=8cm]{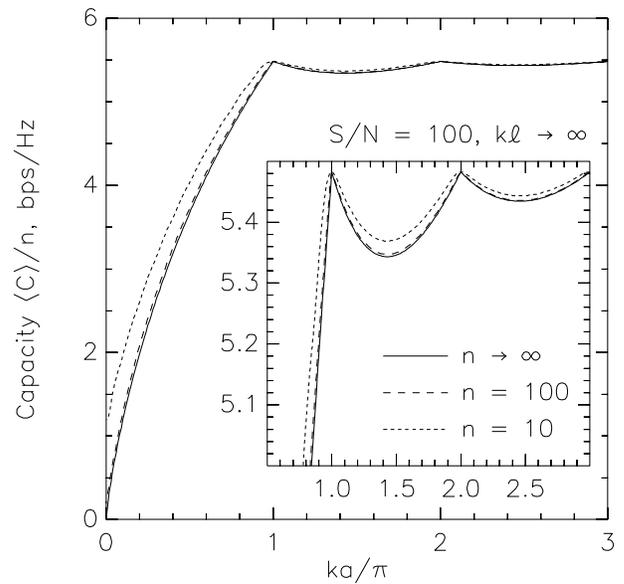}
\caption{\label{fig4} Average information capacity per receiver
of a communication channel
between two identical linear arrays of $n$ equally-spaced transmitters/receivers
as a function of normalized receiver spacing for $k \ell \rightarrow \infty$
and $n \rightarrow \infty$ (solid line), $n = 100$ (dashed line), and
$n = 10$ (dotted line). The signal to noise ratio is $S/N = 100$.
The inset is a zoom of the main plot.}
\end{figure}

We first consider the limit $k \ell \rightarrow \infty$. Although particularly simple
results can be obtained in this case, we will show later that this limit can serve
as a good approximation to real situations with large but finite $k \ell$.
If $0 < k a < \pi$, we find (see Appendix \ref{appB})
\begin{eqnarray}
\frac{\left< C \right>}{n} = \frac{k a}{\pi} \log \left[
\frac{S}{N} \frac{\pi}{ka} \frac{1}{\phi} \right] - \frac{\phi}{\ln2},
\label{c2}
\end{eqnarray}
where
\begin{eqnarray}
\phi = \frac{ka}{\pi} - \frac{1}{2 (S/N)} \left(
\frac{ka}{\pi} \right)^3 \left[ \sqrt{1 + 4 \frac{S}{N} \left(
\frac{\pi}{ka} \right)^2} - 1 \right]. \hspace{5mm}
\label{phi2}
\end{eqnarray}
Analytic results for $\left< C \right>/n$ can also be obtained at $ka > \pi$,
but the resulting expressions are rather cumbersome and lengthy and we present
them in Appendix \ref{appB}.
In Fig.\ \ref{fig4} we show the average capacity per receiver
$\left< C \right>/n$, as a function of normalized receiver spacing $ka/\pi$ at
a fixed signal to noise ratio $S/N = 100$.
The solid line shows our analytic result, corresponding to the limit
$n \rightarrow \infty$ [Eq.\ (\ref{c2}) at $0< ka < \pi$ and more lengthy
but analytic formulas given in Appendix \ref{appB} at $ka > \pi$], while the dashed
and dotted lines are obtained by a numerical solution
of Eqs.\ (\ref{cap})--(\ref{power}) at finite number of receivers $n$ ($n = 100$ and
$10$, respectively). As follows from our analysis, at $k \ell \rightarrow \infty$
and $n \rightarrow \infty$, the derivative of $\left< C \right>$ with
respect to $ka$ exhibits a discontinuity at $ka = m \pi$ ($m = 1, 2, \ldots$).
Although the discontinuity disappears at finite $n$, the overall agreement
between the results corresponding to finite $n$ and $n \rightarrow \infty$ remains
satisfactory even for $n$ as small as
$n = 10$. The analytic results corresponding to the
limit of $n \rightarrow \infty$ can therefore serve as a reasonable approximation
in real situations with large but finite $n$.

\subsubsection{Average capacity at $n \rightarrow \infty$ and
finite $k \ell$}

\begin{figure}
\includegraphics[width=8cm]{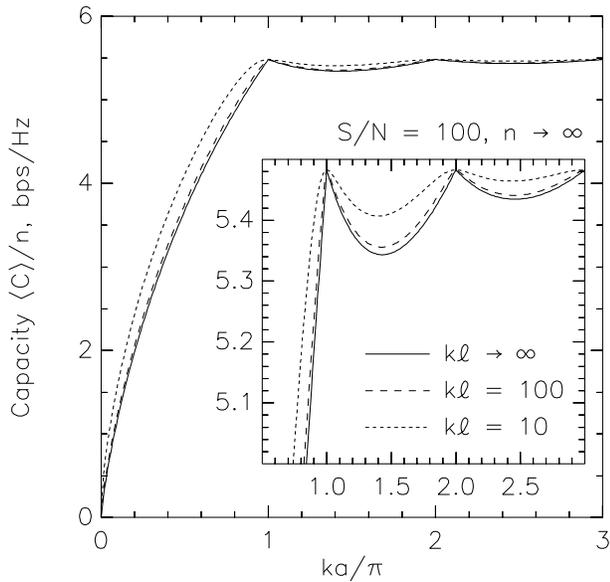}
\caption{\label{fig5} Average information capacity per receiver
of a communication channel
between two identical linear arrays of $n$ equally-spaced transmitters/receivers
as a function of normalized receiver spacing for $n \rightarrow \infty$
and $k \ell \rightarrow \infty$ (solid line), $k \ell = 100$ (dashed line), and
$k \ell = 10$ (dotted line). The signal to noise ratio is $S/N = 100$.
The inset is a zoom of the main plot.}
\end{figure}

Although $k  \ell$ is large in the experiments performed in the diffusion regime
\cite{sebbah00,derode01,fink97,blast}, its value remains
finite and it is therefore of interest to consider its effect on the average
capacity. At finite $k \ell$ the power spectral density $f(\mu)$ is given by
Eq.\ (\ref{f2}) and Eqs.\ (\ref{u1})--(\ref{c1}) can be solved only numerically.
The solution is, however, quite simplified by the fact that the result corresponding
to $k \ell \rightarrow \infty$ is known analytically (see Appendix \ref{appB})
and can be used as a good starting
point for the numerical algorithm. The average capacity per
receiver obtained from Eqs.\ (\ref{u1})--(\ref{c1}) at $k \ell = 100$ and $10$
is shown in Fig.\ \ref{fig5} by dashed and dotted lines, respectively.
The solid line shows the $k \ell \rightarrow \infty$ result, the same as in Fig.\
\ref{fig4}. As follows from Fig.\ \ref{fig5}, at finite $k \ell$
the capacity is somewhat higher than at $k \ell \rightarrow \infty$.
This is explained by a lower degree of correlation between the entries of the
Green matrix $G$ for smaller $k \ell$ [see Eq.\ (\ref{kij})].
 Also,
the derivative of $\left< C \right>$ with respect to $ka$ exhibits no
jumps at $ka = m \pi$ ($m = 1, 2, \ldots$) when $k \ell$ is finite.
In general, however, the average capacity is only slightly affected by the
finiteness of $k \ell$ as long as $k \ell$ remains much larger than unity.

\subsection{Maximum capacity and optimal receiver spacing}

\begin{figure}
\includegraphics[width=8cm]{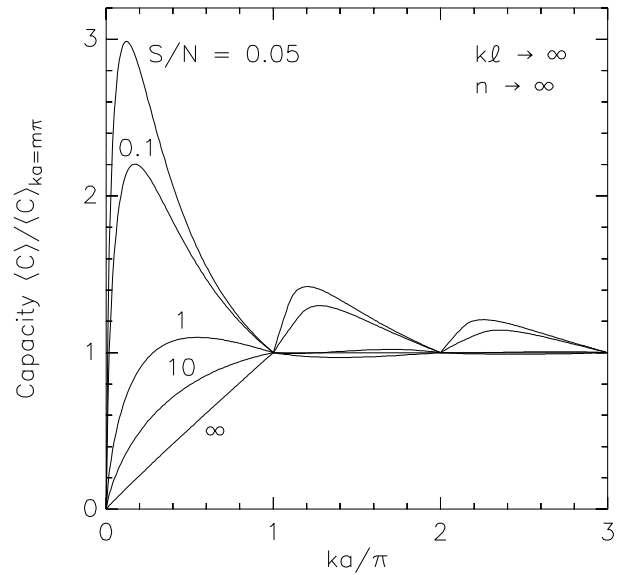}
\caption{\label{fig6} Average capacity of a communication channel between two
identical linear arrays of $n$ transmitters/receivers, normalized to its value
at $ka = m \pi$, is shown as a function of the normalized receiver spacing
for five different values of the signal to noise ratio $S/N$.
We assume $k \ell \rightarrow \infty$ and $n \rightarrow \infty$ for this plot.}
\end{figure}

It follows from Figs.\ \ref{fig4} and \ref{fig5}
that as long as $n \gg 1$ and $k \ell \gg 1$, the average capacity per receiver
is very close
to its value for $n \rightarrow \infty$ and $k \ell \rightarrow \infty$.
We therefore limit the rest of this subsection to the discussion of the latter
limiting case, assuming that the behavior of $\left< C \right>/n$ remains
similar at finite but large $n$ and $k \ell$.
The typical behavior of the average capacity per receiver shown
in Figs.\ \ref{fig4} and \ref{fig5} can be summarized as follows:
$\left< C \right>/n$ has its absolute minimum at $ka \rightarrow 0$ while
it reaches its maxima at $ka = m \pi$ ($m = 1, 2, \ldots$) and becomes almost
independent of $ka$ for $ka > \pi$.
It appears, however, that such a behavior is typical only for \textit{large}
signal to noise ratios $S/N > (S/N)_{c}$, where
$(S/N)_{\mathrm{c}}$ is some critical value of $S/N$ that
we define below.
To study the capacity as a function of $ka$ at various values of $S/N$,
we plot the capacity normalized to its value at $ka = m \pi$ in
Fig.\ \ref{fig6}.

As follows from Fig.\ \ref{fig6}, at large $S/N$
($S/N = 10$ and $S/N \rightarrow \infty$) the behavior of capacity with $ka$ is
similar to that shown in Figs.\ \ref{fig4} and \ref{fig5}. Interestingly,
in the limit $S/N \rightarrow \infty$ we find a very simple result:
$\left< C \right>/n = (ka/\pi) \log(S/N)$ at $ka < \pi$ and
$\left< C \right>/n = \log(S/N)$ at $ka > \pi$, i.e. the capacity grows linearly with
$ka$ for $ka < \pi$ and then remains constant for $ka > \pi$. At finite but
large $S/N$ the behavior of capacity is less simple but is qualitatively very
similar: $\left< C \right>/n$ first shows a monotonic increase with $ka$
for $ka < \pi$ and then oscillates weakly with $ka$ for $ka > \pi$
(see also Figs.\ \ref{fig4} and \ref{fig5}).
As we noted above, such a behavior is typical only for $S/N > (S/N)_{\mathrm{c}}$.
At smaller values of the signal to noise ratio, $\left< C \right>/n$ exhibits
a non-monotonic behavior with $ka$ for $ka < \pi$. More precisely, it reaches
a maximum at some $ka < \pi$, as can be seen from Fig.\ \ref{fig6}
for $S/N = 0.05$, $0.1$, and $1$.
We call the value of $ka$ maximizing the average information capacity
\textit{optimal} and denote it by $(ka)_{\mathrm{opt}}$.
In addition, we define the \textit{maximum} capacity
$\left< C \right>_{\mathrm{max}}$ as the capacity maximized over $ka$:
\begin{eqnarray}
\left< C \right>_{\mathrm{max}} = \max_{ka} \max_{Q}
\left< \log \det \left[ I_n + \frac{S}{N} {\cal G}^+ Q {\cal G} \right] \right>.
\label{capmax}
\end{eqnarray}
It follows from Eq.\ (\ref{c2}) that $(ka)_{\mathrm{opt}}$ and
$\left< C \right>_{\mathrm{max}}$ depend on the signal to noise ratio $S/N$
in a very simple way:
\begin{eqnarray}
(ka)_{\mathrm{opt}} &=&
\cases{
A_1 \sqrt{S/N}, &$S/N < (S/N)_{\mathrm{c}}$,\cr
m \pi, &$S/N > (S/N)_{\mathrm{c}}$,
}
\label{kaopt}
\\
\frac{\left< C \right>_{\mathrm{max}}}{n} &=&
\cases{
A_2 \sqrt{S/N}, \hspace{1cm} S/N < (S/N)_{\mathrm{c}},\cr
\log\left[ (S/N)/\phi_{\mathrm{max}} \right] -
\phi_{\mathrm{max}}/\ln2, \cr
\hspace{2.6cm} S/N > (S/N)_{\mathrm{c}},
}
\label{capmax2}
\end{eqnarray}
where $(S/N)_{\mathrm{c}} \simeq 3.35$, $A_1 \simeq 1.72$ and $A_2 \simeq 0.92$
are numerical constants, and
\begin{eqnarray}
\phi_{\mathrm{max}} =
1 - \frac{1}{2 S/N} \left( \sqrt{1+ 4 S/N} - 1 \right).
\label{phimax}
\end{eqnarray}

\begin{figure}
\includegraphics[width=8cm]{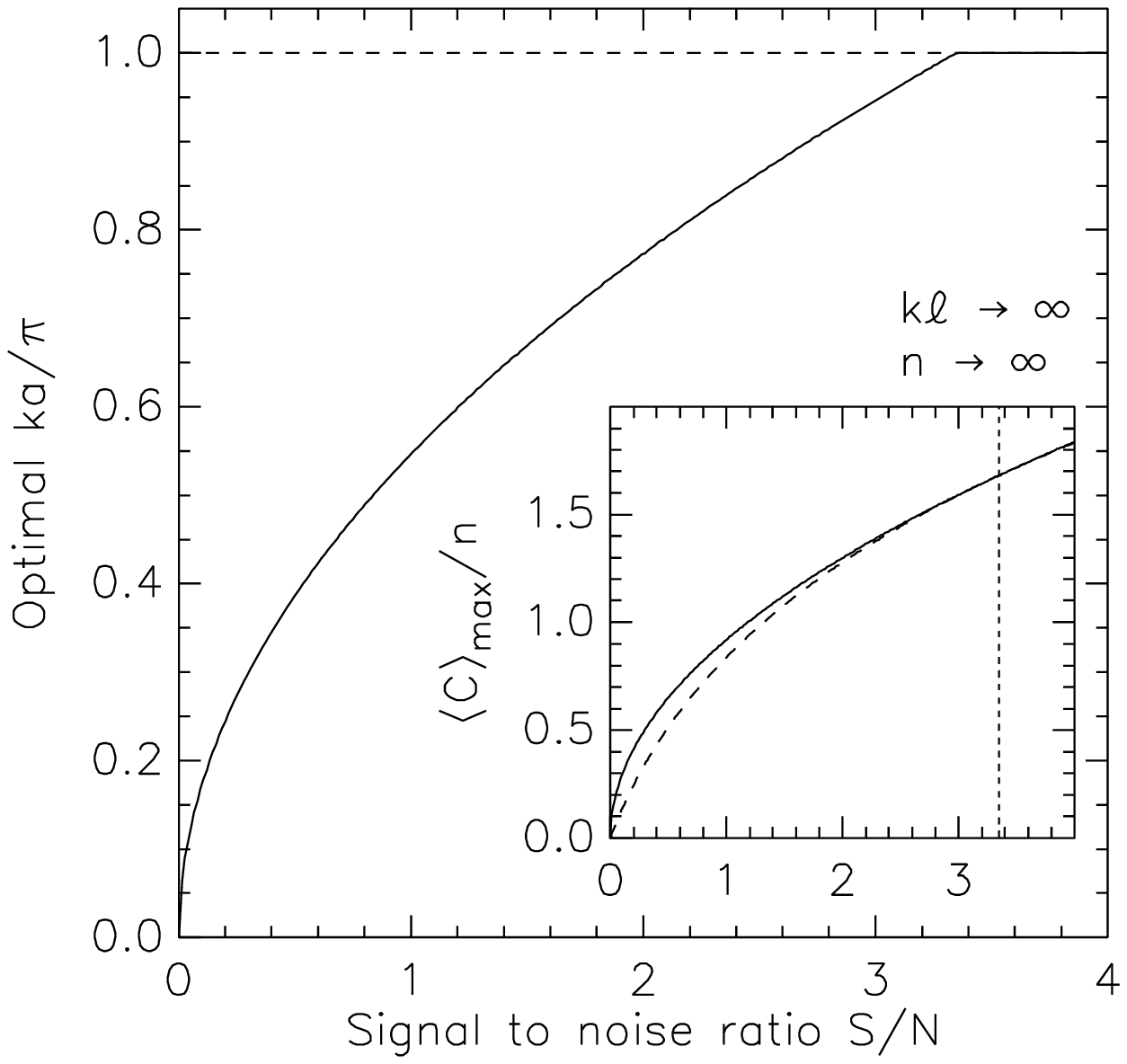}
\caption{\label{fig7}
Normalized optimal receiver spacing, maximizing the information capacity,
as a function of the signal to noise ratio. The inset shows the maximum
average information capacity per receiver $\left< C \right>_{\mathrm{max}}/n$.
The dashed line is $\left< C \right>/n$ at $ka = m \pi$ ($m = 1, 2, \ldots$),
coinciding with $\left< C \right>_{\mathrm{max}}/n$ for
$S/N > (S/N)_{\mathrm{c}} \simeq 3.35$ but smaller than the latter for
$S/N < (S/N)_{\mathrm{c}}$. The vertical dotted line is $S/N = (S/N)_{\mathrm{c}}$.}
\end{figure}

The normalized optimal receiver spacing $(ka)_{\mathrm{opt}}$ is
shown in Fig.\ \ref{fig7},
while the maximum capacity $\left< C \right>_{\mathrm{max}}$ is shown, in the inset.
For comparison, we also show the capacity corresponding to $ka = m \pi$ (the
dashed line in the inset of Fig.\ \ref{fig7}) which coincides with the maximum
capacity for $S/N > (S/N)_{\mathrm{c}}$ but is smaller than the latter if
$S/N < (S/N)_{\mathrm{c}}$. It is interesting to note that at $ka = m \pi$
the average capacity scales linearly with $S/N$ for $S/N \rightarrow 0$,
while the average capacity maximized over $ka$ is proportional to the square root of
$S/N$ in the same limit. Hence, the latter can exceed the former significantly
for small signal to noise ratios.

Qualitatively different behavior of the average capacity at small and large
signal to noise ratios, illustrated in Figs.\ \ref{fig6} and \ref{fig7},
can be understood without any lengthy calculations.
We first remind that at $ka < \pi$, according to Eq.\ (\ref{kij}), smaller
$ka$ means stronger correlation between the Green functions $G_{\alpha i}$.
Next, we map the quite complicated communication channel shown in
Fig.\ \ref{fig1}
onto an equivalent
set of $n^{\,\prime}$ independent communication channels,
the capacity of each equivalent channel being
$\sim \log(1 + S^{\,\prime}/N^{\,\prime})$, where $S^{\,\prime}$ and
$N^{\,\prime}$ denote the signal and noise powers
in each of $n^{\,\prime}$ independent channels. Obviously, the number
$n^{\,\prime}$ of equivalent independent channels grows with decreasing the correlations
between the entries of the Green matrix $G$. Consequently, since
the total capacity is $n^{\,\prime} \log(1 + S^{\,\prime}/N^{\,\prime})$,
it seems that having zero correlations (and hence $ka = m \pi$)
should always maximize the capacity because it ensures the largest $n^{\,\prime}$.
This reasoning, however, is not correct, since the signal power
$S^{\,\prime}$ is also sensitive to the correlations between the
entries of the Green matrix $G$. Indeed, partially correlated Green functions
$G_{\alpha i}$ lead to a constructive interference of scattered waves at
the receivers, thus increasing the power of the received signal, while the noise power
$N^{\,\prime}$ remains unchanged. Therefore, when changing the receiver spacing
$ka$ from $0$ to $\pi$, one gradually switches from a small number
of equivalent independent channels $n^{\,\prime}$ with relatively
large signal to noise ratio $S^{\,\prime}/N^{\,\prime}$
to a larger number of equivalent independent channels
$n^{\,\prime}$ with weaker $S^{\,\prime}/N^{\,\prime}$.
If $S^{\,\prime}/N^{\,\prime}$ is large (which is only possible if
$S/N$ is large), the capacity of each independent channel
$\log(1 + S^{\,\prime}/N^{\,\prime}) \approx
\log(S^{\,\prime}/N^{\,\prime})$ depends on $S^{\,\prime}/N^{\,\prime}$ only
logarithmically, while the dependence of the total capacity on $n^{\,\prime}$
is linear. To achieve the maximum capacity one therefore needs to choose
the largest $n^{\,\prime} = n^{\,\prime}_{\mathrm{max}}$
(which corresponds to $ka = m \pi$), the
decrease of capacity of each equivalent channel being negligible due to its
weak (logarithmic) dependence on the signal to noise ratio.
In contrast, if $S^{\,\prime}/N^{\,\prime}$ is small (which corresponds to small
$S/N$), the capacity of each equivalent channel
$\log(1 + S^{\,\prime}/N^{\,\prime}) \approx
S^{\,\prime}/N^{\,\prime}$ depends linearly on the signal to noise ratio.
In this case, the maximum capacity is achieved by choosing some optimal
number of independent equivalent channels
$n^{\,\prime} < n^{\,\prime}_{\mathrm{max}}$ (and hence some
optimal value of $ka < \pi$), which is large enough, but less than
$n^{\,\prime}_{\mathrm{max}}$
to ensure a reasonable value of $S^{\,\prime}/N^{\,\prime}$ in each of
$n^{\,\prime}$ equivalent channels. This explains the origin of the optimal
receiver spacing and its behavior shown in Fig.\ \ref{fig7}.

\section{Conclusion}
\label{concl}

In the present paper we study the information content of coherent
multiple-scattered wave
fields in disordered media and the capacity of multiple-scattered waves to
transfer the information through a disordered medium.
We show that the information-theoretic quantities, such as the
differential entropy $h$ of the multiple-scattered wave field and the
information capacity $C$
of a communication channel in a disordered medium, are directly related to the
mesoscopic correlations between the scattered waves.
Mesoscopic correlations reduce the information content of the coherent,
multiple-scattered wave field in a disordered medium, leading to a smaller
number of bits required to store all the relevant information about it.
To consider the transfer of information by multiple-scattered waves,
we limit ourselves to the case of communication
between two identical linear arrays of $n$ equally-spaced transmitters/receivers
(receiver spacing $a$).
The average information capacity $\left< C \right>$ of such a communication channel
is shown to scale linearly with $n$, analytic expressions for $\left< C \right>/n$
are obtained in the limit $n \rightarrow \infty$ and $k \ell \rightarrow \infty$.
For finite but large values of $n$ and $k \ell$ the capacity
per receiver $\left< C \right>/n$ is somewhat
greater than at $n \rightarrow \infty$, $k \ell \rightarrow \infty$, but the
latter limiting case proves to be a fairly good approximation as long as
$n \gg 1$ and $k \ell \gg 1$.
Our analysis shows that if the signal to noise ratio $S/N$ exceeds the
critical value $(S/N)_{\mathrm{c}} \simeq 3.35$, $\left< C \right>/n$ grows
monotonically with $a$ as long as $ka <  \pi$ and then oscillates slightly
below its maximum value achieved at $ka = m \pi$ ($m = 1, 2, \ldots$).
If $S/N < (S/N)_{\mathrm{c}}$, the behavior of the average capacity per receiver
$\left< C \right>/n$ is not monotonic for $0 < ka < \pi$ and an absolute
maximum of $\left< C \right>/n$ is reached at some $ka < \pi$.
We define the maximum average capacity $\left< C \right>_{\mathrm{max}}$
as the average capacity maximized over the receiver spacing $a$ and
the normalized optimal receiver spacing $(ka)_{\mathrm{opt}}$ as the
spacing maximizing the average capacity. Both $\left< C \right>_{\mathrm{max}}/n$
and $(ka)_{\mathrm{opt}}$ are proportional to
$(S/N)^{1/2}$ for $S/N < (S/N)_{\mathrm{c}}$.
At $S/N > (S/N)_{\mathrm{c}}$, we find $(ka)_{\mathrm{opt}} = m \pi$ and
$\left< C \right>_{\mathrm{max}}/n \propto \log(S/N)$.

\begin{acknowledgments}
The author is grateful to R. Maynard and B.A. van Tiggelen for numerous discussions.
A.M. Sengupta is acknowledged for useful comments on the preprint \cite{skip02}.

\end{acknowledgments}

\appendix

\section{Average capacity at $n \gg 1$}
\label{appA}

In this Appendix we follow Refs.\ \onlinecite{mous00} and
\onlinecite{sengupta01} to calculate the average capacity $\left< C \right>$
in the large-$n$ limit.
The idea of the calculation stems from the fact that
the moment generating function of the random variable
${\cal C} = \ln \det [ I_n + (S/N) {\cal G}^+ Q {\cal G} ]$,
$F(\gamma) = \left< \exp(-\gamma {\cal C}) \right>$, writes
$F(\gamma) \simeq \exp(-\gamma \left< {\cal C} \right>)$ in the limit of $\gamma
\rightarrow 0^+$. We start therefore by a calculation of $F(\gamma)$,
keeping in mind that taking the limit $\gamma \rightarrow 0^+$ will
allow us to obtain the average capacity $\left< C \right> =
\left< {\cal C} \right>/\ln2$.
We admit that
$F(\gamma) = \left< \left[ \det(I_n + (S/N){\cal G}^+ Q {\cal G})
\right]^{-\gamma} \right>$, and that for integer $\gamma$ we can represent
$\left[ \det(I_n + (S/N) {\cal G}^+ Q {\cal G}) \right]^{-\gamma}$ as
\begin{eqnarray}
&&\left[ \det(I_n + (S/N) {\cal G}^+ Q {\cal G}) \right]^{-\gamma} =
(S/N)^{n \gamma} \int \mathrm{d} \mu(\vec{X}, \vec{Y})
\nonumber \\
&&\times \exp\left[ -\frac{1}{2} (S/N)^{-1/2}
\sum\limits_{m=1}^{\gamma}
\left( \vec{X}_m^+ \vec{X}_m  + \vec{Y}_m^+ \vec{Y}_m \right) \right.
\nonumber \\
&&-\left.
\frac{1}{2}
\sum\limits_{m=1}^{\gamma}
\left( \vec{Y}_m^+ {\cal G}^{\,\prime} \vec{X}_m
- \vec{X}_m^+ {\cal G}^{\,\prime +} \vec{Y}_m \right) \right],
\label{trick1}
\end{eqnarray}
where ${\cal G}^{\,\prime} = Q^{1/2} {\cal G} $ and
we introduce $2 \gamma$ auxiliary complex vectors $\vec{X}_m$ and
$\vec{Y}_m$ ($m = 1, \ldots, \gamma$) --- the procedure known as the ``replica trick'',
--- and $\mathrm{d} \mu(\vec{X}, \vec{Y})$ is the appropriate integration measure.
We now average Eq.\ (\ref{trick1}) over disorder, interchanging the order of averaging
and integration on the right-hand side of the equation:
\begin{eqnarray}
&&F(\gamma) =
(S/N)^{n \gamma} \int \mathrm{d} \mu(\vec{X}, \vec{Y})
\nonumber \\
&&\times \exp\left[ -\frac{1}{2} (S/N)^{-1/2}
\sum\limits_{m=1}^{\gamma}
\left( \vec{X}_m^+ \vec{X}_m  + \vec{Y}_m^+ \vec{Y}_m \right) \right.
\nonumber \\
&&-\left.
\frac{1}{4n}
\sum\limits_{m,l=1}^{\gamma}
\left( \vec{Y}_m^+ A \vec{Y}_l
\vec{X}_l^+ B \vec{X}_m \right) \right],
\label{trick2}
\end{eqnarray}
where $\left< {\cal G}_{\alpha i}^{\,\prime} {\cal G}_{\beta j}^{\,\prime *}
\right> = (1/n) A_{\alpha \beta} B_{i j}$.
The last term in Eq.\ (\ref{trick2}) precludes the direct integration over
$\vec{X}$ and $\vec{Y}$. We can, however, reduce it to a sum of
quadratic terms by defining
$\gamma \times \gamma$ complex matrices $U$ and $V$ and introducing integrations along
appropriate contours in the complex plain:
\begin{eqnarray}
F(\gamma) &=&
(S/N)^{n \gamma} \int \mathrm{d} \mu(\vec{X}, \vec{Y})
\mathrm{d} \mu(U) \mathrm{d} \mu(V)
\nonumber \\
&\times& \exp\left( -\frac{1}{2} {\cal S} \right),
\label{trick3}
\end{eqnarray}
where
\begin{eqnarray}
{\cal S} &=&
(S/N)^{-1/2} \sum\limits_{m=1}^{\gamma}
\left( \vec{X}_m^+ \vec{X}_m  + \vec{Y}_m^+ \vec{Y}_m \right)
\nonumber \\
&+& \sum\limits_{m,l=1}^{\gamma}
\left( \vec{Y}_m^+ A \vec{Y}_l U_{ml} +
V_{ml} \vec{X}_m^+ B \vec{X}_l \right.
\nonumber \\
&-& \left. n U_{ml} V_{lm} \right).
\label{trick4}
\end{eqnarray}
We now perform the integrals over $\vec{X}$ and $\vec{Y}$ in Eq.\ (\ref{trick3})
and are left with
\begin{eqnarray}
&&F(\gamma) =
(S/N)^{n \gamma} \int \mathrm{d} \mu(U) \mathrm{d} \mu(V)
\nonumber \\
&&\times\exp\left\{ -\ln \det \left[ (S/N)^{-1/2} + A U \right] \right.
\nonumber \\
&&-\left.
\ln \det \left[ (S/N)^{-1/2} + B V \right] + n \mathrm{Tr} (U V) \right\},
\label{trick5}
\end{eqnarray}
where $AU$ and $BV$ are the outer products of matrices.

Introducing the eigenvalues $\xi_i$ and $\eta_i$ of the matrices
$A$ and $B$, respectively, we can rewrite the exponent in Eq.\ (\ref{trick5}) as
\begin{eqnarray}
&&\sum\limits_{i=1}^{n} \left\{
\ln \det \left[(S/N)^{-1/2} + \xi_i U \right] \right.
\nonumber \\
&&+ \left.
\ln \det \left[(S/N)^{-1/2} + \eta_i V \right] \right\}
- n \mathrm{Tr}(U V).
\label{trick6}
\end{eqnarray}
In the limit $n \rightarrow \infty$ the integrations in Eq.\ (\ref{trick5}) can
be performed using the saddle point method. Assuming that the replica symmetry
is not broken at the saddle point, we have $U = u I_{\gamma}$ and
$V = v I_{\gamma}$ with $I_{\gamma}$ the $\gamma \times \gamma$ unit matrix.
Equation (\ref{trick6}) then becomes
\begin{eqnarray}
&&\mu \sum\limits_{i=1}^{n} \left\{
\ln \left[(S/N)^{-1/2} + \xi_i u \right] \right.
\nonumber \\
&&+ \left. \ln \left[(S/N)^{-1/2} + \eta_i v \right] \right\}
- \mu n u v.
\label{trick7}
\end{eqnarray}
At the saddle point the partial derivatives of Eq.\ (\ref{trick7}) with respect
to $u$ and $v$ should be zero.
This yields the following equations for $u$ and $v$:
\begin{eqnarray}
u &=& \frac{1}{n} \sum\limits_{i=1}^n \frac{\eta_i}{(S/N)^{-1/2} + \eta_i v},
\label{utrick}
\\
v &=& \frac{1}{n} \sum\limits_{i=1}^n \frac{\xi_i}{(S/N)^{-1/2} + \xi_i u}.
\label{vtrick}
\end{eqnarray}
Relaxing the condition of integer $\gamma$, we take the limit
$\gamma \rightarrow 0^+$ and put Eq.\ (\ref{trick5}) in the form
$F(\gamma) \simeq \exp(-\gamma \left< {\cal C} \right>)$ with
\begin{eqnarray}
\left< {\cal C} \right> &=& \sum\limits_{i=1}^{n} \left\{
\ln \left[(S/N)^{-1/2} + \xi_i u \right] \right.
\nonumber \\
&+& \left.
\ln \left[(S/N)^{-1/2} + \eta_i v \right] \right\}
\nonumber \\
&-& n u v + n \ln(S/N).
\label{trick8}
\end{eqnarray}
The variance of capacity can also be found in a similar way
(see Ref.\ \onlinecite{sengupta01} for details),
but we do not consider it here.
Changing variables back from ${\cal G}^{\,\prime}$ to ${\cal G}$ we see that
$A = Q^{1/2} K Q^{1/2}$ and $B = K$ yielding $\xi_i = \kappa_i q_i$ and
$\eta_i = \kappa_i$ with $\kappa_i$ and $q_i$ the eigenvalues of
the matrices $K$ and $Q$, respectively. Eqs.\ (\ref{utrick})--(\ref{trick8})
then reduce to Eqs.\ (\ref{cap})--(\ref{v}) of Sec.\ \ref{capacitysec}.

To maximize the average capacity $\left< C \right>$ over the ensemble
of covariance matrices $Q$, we consider an infinitesimal
variation $\delta Q$ of the maximizing matrix $Q$ and require
$\delta \left< C \right> \leq 0$.
This leads to $\mathrm{Tr} ( \Phi \delta Q ) \leq 0$, where
\begin{eqnarray}
\Phi = \left< {\cal G} \left[ I_n + (S/N) {\cal G}^+ Q {\cal G}
\right]^{-1} {\cal G}^+ \right>.
\label{phitrick}
\end{eqnarray}
The allowed variations $\delta Q$ of the covariance matrix should keep
$Q + \delta Q$ positive definite and should not change the total emitted power:
$\mathrm{Tr} \delta Q = 0$. If all eigenvalues of $Q$ are positive, the same is true
for $Q + \delta Q$ (provided that $\delta Q$ is small), and
$\mathrm{Tr}( \Phi \delta Q ) = 0$ is achieved by $\Phi = \phi I_n$,
where $\phi$ is a scalar: $\mathrm{Tr}( \Phi \delta Q ) =
\phi \mathrm{Tr} \delta Q = 0$. This can be also shown to remain true if
some eigenvalues of $Q$ are zero \cite{sengupta01}.
It then follows from Eq.\ (\ref{phitrick})
that for $p$ nonzero eigenvalues $q_i$ of the matrix $Q$ one has Eq.\ (\ref{phi})
of Sec.\ \ref{capacitysec}. Finally, Eq.\ (\ref{power}) is simply the
total emitted power constraint $\mathrm{Tr} Q = n$.

\section{Average capacity at
$n \rightarrow \infty$ and $k \ell \rightarrow \infty$}
\label{appB}

In this Appendix we derive analytic expressions for the average information
capacity in the limit of $n \rightarrow \infty$ and $k \ell \rightarrow \infty$.
We consider separately the cases of $0 < ka < \pi$, $\pi < ka <  2\pi$, and
$2\pi < ka <  3\pi$. At $ka > 3 \pi$ calculations can be performed in a similar way.

\subsection{$0 < ka < \pi$}
\label{appB1}

If $0 < ka < \pi$, the power spectral density takes a particularly simple form
[see Eq.\ (\ref{f3})]:
$f(\mu) = \pi/(ka)$ for
$0 < \mu < ka$ or $2 \pi - ka < \mu < 2 \pi$, while $f(\mu) = 0$ for
$ka < \mu < 2 \pi - ka$. Integrations in Eqs.\ (\ref{u1}) and (\ref{v1}) can now
be easily performed, since $f(\mu) > (S/N)^{-1/2} v$ at
$0 < \mu < ka$ or $2 \pi - ka < \mu < 2 \pi$ and
$f(\mu) = 0 < (S/N)^{-1/2} v$ at $ka < \mu < 2 \pi - ka$. This yields
\begin{eqnarray}
u &=& \left[ \left( \frac{S}{N} \right)^{-1/2} + v \frac{\pi}{ka} \right]^{-1},
\label{bu1}
\\
v &=& \frac{ka}{\pi} \frac{1}{u} \left[
1 - \frac{ka}{\pi} v \left( \frac{S}{N} \right)^{-1/2} \right].
\label{bv1}
\end{eqnarray}
We now solve these equations with respect to $u$, $v$ and
substitute the solution into Eq.\ (\ref{c1}), where, again, the integrations
are readily performed. This gives
\begin{eqnarray}
\frac{\left< C \right>}{n} = \frac{k a}{\pi} \log \left[
\frac{S}{N} \frac{\pi}{ka} \frac{1}{\phi} \right] - \frac{\phi}{\ln 2},
\label{bc1}
\end{eqnarray}
where
\begin{eqnarray}
\phi &=& \frac{ka}{\pi} - \frac{1}{2 (S/N)} \left(
\frac{ka}{\pi} \right)^3
\nonumber \\
&\times& \left[ \sqrt{1 + 4 \frac{S}{N} \left(
\frac{\pi}{ka} \right)^2} - 1 \right],
\label{bphi1}
\end{eqnarray}
which are Eqs.\ (\ref{c2}) and (\ref{phi2}) of Sec.\ \ref{capacitysec}.

\subsection{$\pi < ka < 2 \pi$}
\label{appB2}

In this case, $f(\mu) = \pi/(ka)$ for $0 < \mu < 2 \pi - ka$ or
$ka < \mu < 2 \pi$ and $f(\mu) = 2 \pi/(ka)$ for $2 \pi - ka < \mu < ka$.
We now should distinguish two cases: (a) $S/N > (S/N)_1$ and (b) $S/N \leq (S/N)_1$,
where
\begin{eqnarray}
&&\left( \frac{S}{N} \right)_1 = \frac{1}{4} \left( \frac{ka}{\pi} \right)^2
\nonumber \\
&&\times
\left\{ \frac{\sqrt{(ka/\pi) [26 - 7 (ka/\pi)] - 15}}{3 - ka/\pi} - 1 \right\}.
\label{sn1}
\end{eqnarray}

In the case (a), $f(\mu) > (S/N)^{-1/2} v$ for all $\mu \in (0, 2 \pi)$
and Eqs.\ (\ref{u1}), (\ref{v1}) become
\begin{eqnarray}
u &=& \frac{2 - ka/\pi}{v + (S/N)^{-1/2} ka/\pi}
\nonumber \\
&+& \frac{ka/\pi - 1}{v + (S/N)^{-1/2} ka/(2 \pi)},
\label{bu2a}
\\
v &=& \frac{1}{u} \left[ 1 - \frac{ka}{2 \pi} \left( \frac{S}{N} \right)^{-1/2} v
\left( 3 - \frac{ka}{\pi} \right) \right].
\label{bv2a}
\end{eqnarray}
The values of $u$ and $v$ found from the two above equations are to be substituted
into Eq.\ (\ref{c1}) that reduces to
\begin{eqnarray}
\frac{\left< C \right>}{n} &=&
\left( 2 - \frac{ka}{\pi} \right) \log \left\{ \frac{\pi}{ka}
\left[ \frac{1}{v (S/N)^{1/2}} + \frac{\pi}{ka} \right] \right\}
\nonumber \\
&+& \left( \frac{ka}{\pi} - 1 \right) \log \left\{ \frac{2 \pi}{ka}
\left[ \frac{1}{v (S/N)^{1/2}} + \frac{2 \pi}{ka} \right] \right\}
\nonumber \\
&-& \frac{u v}{\ln 2} + \log\frac{S}{N}.
\label{bc2a}
\end{eqnarray}

In the case (b), $f(\mu) > (S/N)^{-1/2} v$ only for
$2 \pi - ka < \mu < ka$. Eq.\ (\ref{bu2a}) remains the same, while
Eqs.\ (\ref{bv2a}) and (\ref{bc2a}) become
\begin{eqnarray}
v &=&\frac{1}{u} \left( \frac{ka}{\pi} - 1 \right)
\left[ 1 - \left( \frac{S}{N} \right)^{-1/2} v \frac{ka}{2 \pi} \right],
\label{bv2b}
\\
\frac{\left< C \right>}{n} &=&
\left( 2 - \frac{ka}{\pi} \right) \log \left\{ \left( \frac{S}{N} \right)^{-1/2}
\left[ \left( \frac{S}{N} \right)^{-1/2} + v \frac{\pi}{ka} \right] \right\}
\nonumber \\
&+& \left( \frac{ka}{\pi} - 1 \right) \log \left\{ \frac{2 \pi}{ka}
\left[ \frac{1}{v (S/N)^{1/2}} + \frac{2 \pi}{ka} \right] \right\}
\nonumber \\
&-& \frac{u v}{\ln 2} + \log\frac{S}{N}.
\label{bc2b}
\end{eqnarray}

\subsection{$2\pi < ka < 3\pi$}
\label{appB3}

We proceed as in the two previous subsections.
$f(\mu) = 3 \pi/(ka)$ for $0 < \mu < ka - 2 \pi$ or
$4 \pi - ka < \mu < 2 \pi$, while
$f(\mu) = 2 \pi/(ka)$ for $ka - 2 \pi < \mu < 4 \pi - ka$.
If $S/N > (S/N)_2$ [case (a)], where
\begin{eqnarray}
&&\left( \frac{S}{N} \right)_2
\nonumber \\
&&= \frac{(ka/\pi)^2 [ka/(2 \pi) - 1 ]}{\sqrt{(ka/\pi) (32 - 5 ka/\pi) - 35}
+ 5 - ka/\pi},\hspace{5mm}
\label{sn2}
\end{eqnarray}
$f(\mu) > (S/N)^{-1/2} v$ for all $\mu \in (0, 2 \pi)$ and
Eqs.\ (\ref{u1})--(\ref{c1}) reduce to
\begin{eqnarray}
u &=& \frac{3 (ka/\pi - 2)}{3 v + (S/N)^{-1/2} ka/\pi}
\nonumber \\
&+& \frac{3 - ka/\pi}{v + (S/N)^{-1/2} ka/(2\pi)},
\label{bu3a}
\\
v &=& \frac{1}{u} \left[ 1 - \frac{ka}{6 \pi} \left( \frac{S}{N} \right)^{-1/2} v
\left( 5 - \frac{ka}{\pi} \right) \right],
\label{bv3a}
\\
\frac{\left< C \right>}{n} &=&
\left( \frac{ka}{\pi} - 2\right) \log \left\{ \frac{3 \pi}{ka}
\left[ \frac{1}{v (S/N)^{1/2}} + \frac{3\pi}{ka} \right] \right\}
\nonumber \\
&+& \left( 3 - \frac{ka}{\pi} \right) \log \left\{ \frac{2 \pi}{ka}
\left[ \frac{1}{v (S/N)^{1/2}} + \frac{2 \pi}{ka} \right] \right\}
\nonumber \\
&-& \frac{u v}{\ln 2} + \log\frac{S}{N}.
\label{bc3a}
\end{eqnarray}

If $S/N < (S/N)_2$ [case (b)],
$f(\mu) > (S/N)^{-1/2} v$ for $0 < \mu < ka - 2 \pi$ or
$4 \pi - ka < \mu < 2 \pi$ only and we have instead of
Eqs.\ (\ref{bv3a}) and (\ref{bc3a})
\begin{eqnarray}
v &=& \frac{2}{u} \left[ \frac{ka}{2 \pi} - 1
- \frac{ka}{3 \pi} \left( \frac{S}{N} \right)^{-1/2} v
\left( \frac{ka}{2\pi} - 1 \right) \right],
\label{bv3b}
\\
\frac{\left< C \right>}{n} &=&
\left( \frac{ka}{\pi} - 2 \right) \log \left\{ \frac{3 \pi}{ka}
\left[ \frac{1}{v (S/N)^{1/2}} + \frac{3\pi}{ka} \right] \right\}
\nonumber \\
&+& \left( 3 - \frac{ka}{\pi} \right) \log \left\{ (S/N)^{-1/2}
\left[ (S/N)^{-1/2} + \frac{2 \pi}{ka} v \right] \right\}
\nonumber \\
&-& \frac{u v}{\ln 2} + \log\frac{S}{N}.
\label{bc3b}
\end{eqnarray}


\end{document}